\documentclass[11pt]{article}
\usepackage{latexsym}
\usepackage{amsmath}
\usepackage{amssymb}
\usepackage{bbm}
\usepackage{graphicx}
\usepackage{subfigure}
\usepackage{dcolumn} 
\usepackage{bm}
\usepackage{booktabs}
\usepackage{cite}
\usepackage{authblk}
\usepackage{comment}
\usepackage[usenames,dvipsnames]{color}
\usepackage[]{hyperref}

\usepackage{tikz}
\usepackage{geometry}

\usetikzlibrary{arrows}
\usepackage{fancyhdr}
\usepackage{xcolor}
\newsavebox\MBox

\usepackage{setspace}
\newcommand{\be}{\begin{equation}}
\newcommand{\ee}{\end{equation}}
\newcommand{\bea}{\begin{eqnarray}}
\newcommand{\eea}{\end{eqnarray}}

\newcommand{\dd}{{\rm d}}

\newcommand{\ZZ}{{\mathbb Z}}

\newcommand{\CC}{{\mathbb C}}
\newcommand{\PP}{{\mathbb P}}
\newcommand{\WW}{{\mathbb W}}

\newcommand{\no}{\nonumber}

\def\vev#1{\left\langle #1 \right\rangle}

\def\l:{\mathopen{:}\,}
\def\r:{\,\mathclose{:}}

\newcommand{\kahler}{{K\"{a}hler }}

\newcommand{\bphi}{\overline{\phi}}
\newcommand{\bpsi}{\overline{\psi}}

\newcommand{\cO}{{\mathcal{O}}}

\title{Notes on GLSMs for Supermanifolds and Their Mirrors}
\author[1,2]{Hao Zou\footnote{haozou@fudan.edu.cn}}
\affil[1]{Center for Mathematics and Interdisciplinary Sciences, \protect\\ Fudan University, Shanghai 200433, China}
\affil[2]{Shanghai Institute for Mathematics and Interdisciplinary Sciences, \protect\\  Shanghai 200433, China}
\date{ \small \today}

\begin{document}
\maketitle
\begin{abstract}
	In this paper, we revisit the A-twisted gauged linear sigma models (GLSMs) whose geometric phases are complex K\"ahler supermanifolds. For abelian models without superpotentials we propose an explicit orbifold description of the non-geometric (Landau-Ginzburg) point, and give a systematic rule for the nontrivial R-charge assignments at that point. We then study topological super Landau-Ginzburg models, derive chiral ring relations and genus-$g$ correlation functions, and use these formulas to test a Hori-Vafa-type mirror proposal for supermanifolds.
\end{abstract}

\newpage
\tableofcontents


\section{Introduction}
\label{sec:introduction}

Supermanifolds are the generalizations of ordinary manifolds by incorporating Grassmannian odd directions. They provide the geometric setting for superspaces that have been extensively used in supersymmetric theories. The gauged linear sigma model (GLSM) construction for complex \kahler supermanifolds was first proposed in \cite{Seki:2005hx}, extending GLSM for ordinary manifolds \cite{Witten:1993yc}. In the previous work \cite{Gu:2018xzx}, we have studied the A-twisted GLSMs for supermanifolds and calculated correlation functions and elliptic genera by supersymmetric localization on Coulomb branch. These GLSMs, serving as the ultraviolet (UV) completions, undergo renormalization group (RG) flow to nonlinear sigma models (NLSMs) on supermanifolds in the geometric phase \cite{Seki:2005hx,Belhaj:2004ts,Gu:2018xzx}. As in the ordinary manifold case \cite{Jockers:2012dk,Gomis:2012wy}, we also assume that this RG flow can be characterized by two-sphere partition function $Z_{S^2}$. Consequently, our results provide an evidence for the equivalence between NLSMs defined on a supermanifold and those on specific hypersurfaces or complete intersections within ordinary geometries \cite{Sethi:1994ch, Schwarz:1995ak}. This equivalence relation had been verified from the mirror side in \cite{Aganagic:2004yh} as well.

For the purpose of establishing the equivalence relation, we were restricted to the geometric phase in \cite{Gu:2018xzx}. The A-twisted GLSMs we have considered all have no superpotentials. In the ordinary manifold case, the other non-geometric phase is usually described by the effective twisted superpotential on Coulomb branch. However, this is not true for supermanifolds. As in the main example $\PP^{4|1}$ we are going to consider in Section~\ref{sec:review}, there would be no Coulomb vacua anymore when $r \ll 0$. Instead, we propose an orbifold description for these non-geometric phases, together with the non-trivial R-charges. More specifically, the $r \ll 0$ phase  of the GLSM for $\PP^{4|1}$ is described by 5 even chiral fields on the orbifold $\CC^5/\ZZ_5$ with R-charges $\frac{2}{5}$. We would like to call this the Landau-Ginzburg point (LG point) of the GLSM for $\PP^{4|1}$, since it can be understood as the zero superpotential limit of a Landau-Ginzburg model defined on $\CC^5/\ZZ_5$ with superpotential $W$ of degree $5$. The whole picture is summarized in Figure~\ref{fig:relation}, including the equivalence relation mentioned above.

The study of GLSM for supermanifolds can be enriched by mirror symmetry. A Hori-Vafa-like mirror was first proposed by \cite{Aganagic:2004yh}, and was further investigated in, for example, \cite{Belhaj:2004ts,Garavuso:2011nz}. In this note, we will also chase this direction and give more evidence for the mirror proposal in Section~\ref{sec:mirror}. We will start with the super Landau-Ginzburg models and derive the formulas for chiral ring relations and correlation functions in the general set-up. Then, we apply these formulas to the mirror model, which is a specific super Landau-Ginzburg model, and match chiral ring relations and correlation functions from both mirror sides. Therefore, the relations among GLSMs and mirror models can be illustrated in Figure~\ref{fig:relation2}.

The main result of this paper is that we complete the diagrams relating GLSMs for supermanifolds, GLSMs for ordinary manifolds as well as their mirror models. One such example is shown in Figure~\ref{fig:relation} and~\ref{fig:relation2}. In particular, we propose the orbifold description for the Landau-Ginzburg point of GLSMs for supermanifolds. After a close look at the super Landau-Ginzburg, we provide more evidence to support the mirror symmetry proposed in \cite{Aganagic:2004yh} by matching chiral ring relations and correlation functions.

The paper is organized as follows. In Section~\ref{sec:review}, we start with the main example, the GLSM for $\PP^{4|1}$, and discuss its low energy behavior at the LG point. In particular, the rule for the non-trivial R-charge assignments at the LG point is presented in Section~\ref{sec:rcharge}. With this, we will give more examples beyond the main example. In Section~\ref{sec:mirror}, we will address on the mirror symmetry of GLSMs for supermanifolds and give more evidence for the mirror proposal. The chiral ring relations and correlation functions for super Landau-Ginzburg models are derived in the general cases in this section. The derivation of partition functions and the verification of mirror symmetry for general cases are summarized in Appendix~\ref{app:corf} and~\ref{app:verf}, respectively.

\section{The LG Point of GLSM for Supermanifolds}
\label{sec:review}
In this section, we mainly discuss the non-geometric phase ($r\ll 0$) of the GLSM for supermanifolds. Inspired from the equivarience between GLSM for supermanifolds and for ordinary submanifolds \cite{Gu:2018xzx}, we propose the orbifold description for this phase, with non-trivial R-charge assignments.

\subsection{GLSM for \texorpdfstring{$\PP^{4|1}$}{TEXT} revisited}
Let us start with $U(1)$ GLSM for $\PP^{4|1}$. Here, we follow the conventions in \cite{Belhaj:2004ts,Seki:2005hx,Gu:2018xzx} and we have five Grassmann even chiral fields $\Phi_i$ with $U(1)$ charges $1$ and one Grassmann odd chiral field $\Psi$ with $U(1)$ charge $5$. Here, the Grassmann even/odd indicates the statistical properties of the first scalar components of the chiral multiplets. In this model, there is no superpotential. 
The classical potential energy for the dynamical scalar fields is given as
\be
\label{eq:scalarpotential}
  U \:=\: \frac{1}{2e^2}D^2 + |\sigma|^2 \sum_i \left( \bphi_i\phi_i + 5^2 \bpsi\psi \right) \,,
\ee
where the auxiliary field $D$ can be eliminated by using its equation of motion
\be
  -\frac{D}{e^2} \:=\: \sum_{i=1}^5 \bphi_i\phi_i + 5 \bpsi\psi - r\, ,
\ee
The classical vacuum is determined by solutions to $U=0$, which further implies the $D$-term equation $D=0$. In the phase of $r \gg 0$, $D=0$ requires that not all $\phi_i$'s and $\psi$ can be zero, and the conclusion is that the vacuum would be the super toric variety $\PP^{4|1}$ \cite{Seki:2005hx,Gu:2018xzx}, which is defined by the equivalent relation in $\CC^{5|1}$:
\be
\label{eq:eqrelation}
	(\phi_1, \cdots, \phi_5, \psi) \sim (\lambda \phi_1, \cdots, \lambda \phi_5, \lambda^5 \psi), \quad \lambda \in \CC^*\,.
\ee
Therefore, this phase is described by the nonlinear sigma model on $\PP^{4|1}$. 

Let us now turn to the phase $r \ll 0$ and explore what this region would look like. In analogy of the discussion from \cite{Witten:1993yc,Witten:1993xi}, there are two possibilities to obtain solutions to the $D$-term equation in $r \ll 0$. The first possibility is to consider the quantum corrections, in analogy to the cases of ordinary projective spaces and Grassmannians. In this situation, we shall consider a nonvanishing vev for the gauge field strength $\sigma \neq 0$, then both even and odd chiral fields become massive due to the couplings in \eqref{eq:scalarpotential}, and they should be all integrated out. Therefore, the low energy effective theory ends on the pure Coulomb branch and it is governed by the one-loop corrected twisted superpotential, or called effective twisted superpotential, which can be computed following \cite{Seki:2005hx}
\be
  \widetilde{W}_{\rm eff}(\Sigma) \:=\: -t \Sigma - 5 \Sigma \left(\ln \Sigma -1 \right) + 5 \Sigma\left(\ln 5\Sigma - 1 \right)\,.
\ee
However, one can immediately see that there is no Coulomb vacua as the equation of motion $\exp(\partial \widetilde{W}_{\rm eff}(\sigma)) = 1$. Therefore, the assumption that this phase only consists of pure Coulomb branch contradicts with the Witten index obtained in the phase $r \gg 0$, which was demonstrated to be the same as the quintic $\PP^4[5]$ in \cite{Gu:2018xzx}. To be more precise, the authors of \cite{Gu:2018xzx} have demonstrated the equivalence of elliptic genera via supersymmetric localization, and this equivalence leads to the same Witten index. 

\begin{figure}[!h]
\centering
\begin{tikzpicture}[every text node part/.style={align=center},>=angle 90]
  \draw (0,0) node (n1) {LG point \\ orbifold $\CC^5/\ZZ_5$};
  \draw (0,-3.5) node (n2) {LG on $\CC^5/\ZZ_5$ w/ \\ superpotential $W$ };
  \path[->, line width=0.3mm, font = \footnotesize ](n2) edge node[draw=none,sloped, anchor=center, above] {$W \longrightarrow 0$} (n1);
  \draw (10,0) node (n3) { NLSM \\on $\CC\PP^{4|1}$};
  \draw (10,-3.5) node (n4) {NLSM \\on quintic};
  \path[<->,line width=0.3mm] (n4) edge node[draw=none,sloped, anchor=center, below]{Equivalent} (n3);
  \draw (5,0) node (n5) {GLSM \\ for $\CC\PP^{4|1}$};
  \draw (5,-3.5) node (n6) {GLSM \\ for quintic};
  \path[<->,line width=0.3mm] (n6) edge node[draw=none,sloped, anchor=center, below]{Equivalent} (n5);
  \path[->,line width=0.3mm] (n5) edge node[above]{$r \ll 0$} (n1);
  \path[->,line width=0.3mm] (n5) edge node[above]{$r \gg 0$} (n3);
  \path[->,line width=0.3mm] (n6) edge node[above]{$r \ll 0$} (n2);
  \path[->,line width=0.3mm] (n6) edge node[above]{$r \gg 0$} (n4);
\end{tikzpicture} 
\caption{Relations between different phases of A-twisted GLSMs.}
\label{fig:relation}
\end{figure}
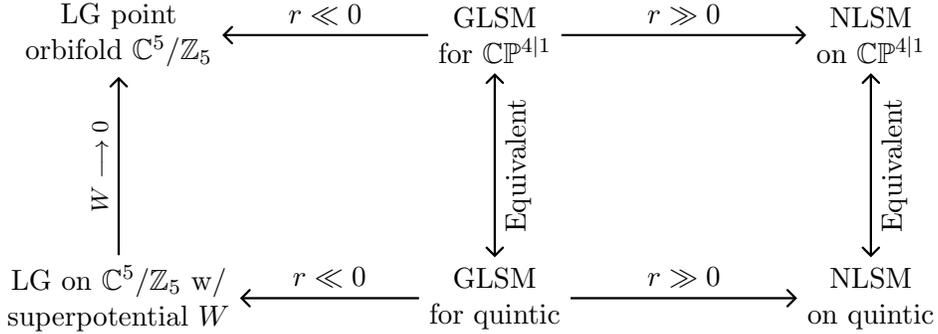

The second possibility is that the phase $r \ll 0$ of this GLSM for $\PP^{4|1}$ consist of a pure Higgs branch. Because $\sum_{i=1}^5 \bphi_i\phi_i $ is always nonnegative, the $D$-term equation requires that the vacuum expectation $\vev{\bpsi\psi}$ \footnote{Since there is no superpotential, we cannot obtain the vev for $\vev \psi $ and then do the perturbation. From the D-term, we can only conclude the vev for $\vev{\bpsi\psi}$. } should be negative when $r \ll 0$. 
Due to the $\CC^*$-action \eqref{eq:eqrelation}, the degree of freedom of $\phi_i$'s are kept while $\vev{\bpsi\psi}$ can be fixed to be a nonzero constant. Considering the low energy behaviors, the Grassmann odd chiral field $\psi$ becomes massive due to nonvanishing value of $\vev{\bpsi\psi}$, and it should be integrated out. Moreover, one can learn more about this phase if we admit the equivalence relation between the GLSMs for $\PP^{4|1}$ and $\PP^4[5]$ as illustrated in \cite{Gu:2018xzx}. We conjecture that this phase is just the orbifold $\CC^5/\ZZ_5$, which can be understood as the zero potential limit of the Landau-Ginzburg orbifold on $\CC^5/\ZZ_5$ as depicted by the up-arrow in Figure~\ref{fig:relation}. Equivalently, this zero superpotential limit can also be viewed as the zero limit of scaling factor, $\lambda \to 0$, as discussed in \cite{Guffin:2008kt}. Therefore, the $U(1)$ gauge also breaks to the discrete symmetry $\ZZ_5$ when $r \ll 0$. From this viewpoint, we will call the phase $r\ll 0$ the {\it Landau-Ginzburg point} of the GLSM for $\PP^{4|1}$, and this is also consistent with the fact that superpotentials are $Q$-exact in A-twisted models \cite{Witten:1993yc}. 

One shall note that the R-charge for the five even chiral fields $\phi_i$ should be nonzero. More specifically, one should get $R_i=\frac{2}{5}$ since taking the zero potential limit does not affect the R-charge assignment. We will give more discussion on the R-charge from a different perspective in the next subsection, providing a supporting evidence. In this sense, the non-vanishing R-charges of those even chiral fields have the same effect of ``compactification'' as the superpotential in A-twisted models. 


Combining the results in \cite{Schwarz:1995ak,Sethi:1994ch,Gu:2018xzx} with the previous discussions, we can find out the relations of A-twisted GLSMs and their low energy effective theories as depicted in Figure~\ref{fig:relation}.

\subsection{R-charges}
\label{sec:rcharge}
The rule for general R-charge assignment was discussed in the Appendix of \cite{Gu:2018xzx}. In this subsection, we want to discuss the rule to assign R-charges in different phases of a GLSM. 

Before discussing about the supermanifold case, let us review the R-charges in A-twisted GLSM for ordinary manifolds. More concretely, let us review the GLSM for quintic, which is equivalent to previous GLSM for $\PP^{4|1}$ as mentioned in \cite{Gu:2018xzx,Aganagic:2004yh}. Although it is necessary to address that GLSM for quintic and GLSM for $\PP^{4|1}$ are different theories as they have different global symmetries \cite{Gu:2018xzx}. GLSM for quintic is defined by \cite{Witten:1993yc}: $5$ chiral fields $\Phi_i$ with $U(1)$ charges $1$ and $1$ chiral field $P$ with $U(1)$ charge $-5$, together with a superpotential $W = PG(\Phi)$, where $G(\Phi)$ is a homogeneous function of $\Phi$ of degree $5$. In phase $r \gg 0$, its low energy physics is described by a nonlinear sigma model on quintic, while in phase $r \ll 0$ it is described by a Landau-Ginzburg model on $\CC^5/\ZZ_5$ with superpotential $G(\Phi)$.

In an A-twisted GLSM, the R-charges assigned to $\Phi_i$'s and $P$ are only constrained to make sure the superpotential $W$ has R-charge $2$ \cite{Witten:1993yc}. However, when it comes to its low energy physics, the R-charges for those chiral fields should be chosen more carefully in different phases to keep the supersymmetry. More specifically, in $r \gg 0$, the vacuum is a quintic, therefore R-charges for $\Phi_i$'s should be zero so that $\phi_i$'s as coordinates for the vacuum are invariant under R-symmetry and while the R-charge for $P$ field should be $2$. In the region where $r \ll 0$, the $P$ field takes expectation value of $\sqrt{-r/5}$, therefore R-charge for the $P$ field should be zero and so R-charges for $\Phi_i$'s are $\frac{2}{5}$.

Now back to the supermanifold case, the R-charges for GLSM for supermanifolds have been discussed in \cite{Gu:2018xzx}. The general result from the discussion there is that the R-charge for even and odd chiral fields should be proportional to their gauge charges. More specifically, we should have the following: in $r\gg 0$, the R-charges for both even and odd chiral fields are zeros; in $r \ll 0$, the R charge for even chiral fields $\Phi_i$ should be $\frac{2}{5}$ while the R-charge for odd chiral field $\Psi$ should be $2$.

The above choice of R-charges is motivated by the following. In $r \gg 0$, the R-charges for both $\Phi_i$ and $\Psi$ should vanish as the vacuum is $\PP^{4|1}$ \cite{Gu:2018xzx}, and only in this way the supersymmetry is kept. The situation of $r \ll 0$ is more subtle. First, note that $\vev{\bpsi\psi} \neq 0$ and the total R-charge of $\bpsi\psi$ should be zero. But the R-charge for $\psi$ can still be nonzero, which is different with the $P$ field in the quintic case. Second, from the CFT perspective, in $r \gg 0$, the central charge can be obtained $\hat{c} = 4 - 1 = 3$ \cite{Seki:2006cj}. Correspondingly, in $r \ll 0$, the central charge should obey $\hat{c} = \sum_{i=1}^5(1-R_i) = 3$, viz., $R_i = \frac{2}{5}$ and then $R_{\psi} = 2$. Here is another argument that we could end up with the same answer but from a different perspective. From the equivalence relation between GLSM for $\PP^{4|1}$ and GLSM for quintic, there is a relation between $\Psi$ and the $P$ field, especially the relation between their R-charges. In $r \gg 0$, $R_p = 2$ and $R_\psi = 2 - R_p = 0$. Then in $r \ll 0$, $R_p = 0$ and so $R_\psi = 2 - R_p = 2$. Therefore, as in the appendix of \cite{Gu:2018xzx}, we can obtain $R_i = \frac{2}{5}$. 

\subsection{Examples beyond \texorpdfstring{$\PP^{4|1}$}{TEXT}}

So far the discussions have been focusing on the typical example $\PP^{4|1}$, and it turns out these discussions can go beyond this example and can be used in more general setting ups. Therefore, we will present more general examples in this section, including super weighted projective spaces and product of superspaces. 

\subsubsection{Super weighted projective spaces}
The generalization to this case is straightforward. Consider $U(1)$ GLSM for supermanifolds with one odd direction but with more general gauge charges, viz., $\WW\PP^{m|1}$. The chiral fields of this model is given as below
\begin{itemize}
  \item $m+1$ even chiral fields, $\Phi_i$, with gauge charge $Q_i$,
  \item $1$ odd chiral fields, $\Psi$, with gauge charge $\tilde{Q}$,
\end{itemize}
where we assume $Q_i$'s and $\tilde{Q}$'s are all positive and they satisfy Calabi-Yau condition:
\be
	\sum_{i=1}^{m+1} Q_i \:=\: \tilde{Q}\,.
\ee
Previous analysis can be directly used in this example. The $D$-term is 
\be
  D \:\sim\: \sum_{i=1}^{m+1} Q_i \bphi_i \phi_i + \tilde{Q}\bpsi \psi - r \,. 
\ee
In the phase $r \gg 0$, the low energy effective theory is a nonlinear sigma model on $\WW\PP^{m|1}$. While in $r \ll 0$, $\vev{\bpsi\psi}$ shall be nonzero and the $U(1)$ gauge breaks down to $\ZZ_{\tilde{Q}}$. We should assign to each even chiral field $\phi_i$ the R-charge $\frac{2 Q_i}{\tilde{Q}}$, while to the odd chiral field $\psi$ the R-charge $2$. Therefore, the Landau-Ginzburg point is described by the orbifold $\CC^m/\ZZ_{\tilde{Q}}$, and there is no superpotential. 

\subsubsection{Multiple Grassmann-odd directions}

We can include more than one Grassmann-odd chiral fields to make the corresponding GLSM to describe more superspaces. For example, GLSM for $\WW\PP^{m|n}$, which is defined by
\begin{itemize}
  \item $m+1$ even chiral fields, $\Phi_i$, with gauge charge $Q_i$,
  \item $n$ odd chiral fields, $\Psi_\mu$, with gauge charge $\tilde{Q}_\mu$,
\end{itemize}
where we assume $Q_i$'s and $\tilde{Q}_{\mu}$'s are all positive and they satisfy Calabi-Yau condition:
\be
    \sum_{i=1}^{m+1} Q_i - \sum_{\mu=1}^{n} \tilde{Q}_\mu \:=\: 0\,.
\ee

The $D$-term in this GLSM is
\be
    D \:\sim\: \sum_{i=1}^{m+1}Q_i \bphi_i\phi_i + \sum_{\mu=1}^{n}\tilde{Q}_\mu \bpsi_\mu \psi_\mu - r\,.
\ee
In $r \gg 0$, the target space of the low energy NLSM is the weighted super toric variety $\WW\PP^{m|n}$ and the weights are given by $Q_i$'s and $\tilde{Q}_{\mu}$'s. However, for $r \ll 0$, story will be more complicated than previous cases. A similar analysis following \cite{Witten:1993yc} can conclude that this phase shall be described by a hybrid sigma/orbifold model, where the orbifold bundle is given by $\CC^m/(\ZZ_{\tilde{Q}_1}\times \cdots \times \ZZ_{\tilde{Q}_n})$. 

Let us look at a simpler example $\PP^{2n-1|2}$ for some integer $n>1$. The $U(1)$ gauge charges for the even and odd chiral fields are given as $Q_i=1$ for $i=1,\dots,2n$ and $\tilde{Q}_1 = \tilde{Q}_2 = n$. The R-charges of $\phi_i$'s and $\psi_\mu$'s for both $r\gg 0$ and $r \ll 0$ are summarized as in Table~\ref{tab:twoodd}.
\begin{table}[h!]
\centering
\renewcommand{\arraystretch}{1.2}
\begin{tabular}[h!]{cccccc}
\toprule
  ~  & $\phi_1$ & $\cdots$ & $\phi_n$ &$\psi_1$ &$\psi_2$ \\\midrule 
  $U(1)$  & 1 &$\cdots$  &1 & n &n \\
  $\left.R\right|_{r\gg 0}$ & 0  &$\cdots$  &0 &0 &0\\
  $\left.R\right|_{r\ll 0}$ & $\frac{2}{n}$  &$\cdots$  &$\frac{2}{n}$ &$2$ &$2$\\
\bottomrule
\end{tabular}
\caption{Charge table for $\PP^{2n-1|2}$.}
\label{tab:twoodd}
\end{table}

The Landau-Ginzburg point of this GLSM is the orbifold $\CC^{2n}/(\ZZ_n\times \ZZ_n)$, and there are two independent discrete gauge transformations acting on $\phi_i$ by multiplying $\exp(2\pi i/n)$.

\subsubsection{Product of superspaces}
A further generalization is to consider the product of superspaces. For example, consider the product space $\WW\PP^{n|1}\times \WW\PP^{m|1}$, which can be realized by a $U(1)^2$ GLSM for $\WW\PP^{n|1}\times \WW\PP^{m|1}$. The charges of the even and odd chiral fields are given in Table~\ref{tab:productgauge}.
\begin{table}[h!]
\centering
\renewcommand{\arraystretch}{1.2}
\begin{tabular}{ccccccccc}
\toprule
  ~  & $\phi_{1,1}$ & $\cdots$ & $\phi_{1,n+1}$ & $\phi_{2,1}$ & $\cdots$ & $\phi_{2,m+1}$ &$\psi_1$ &$\psi_2$ \\\midrule 
  $U(1)_1$  & $Q^1_1$ &$\cdots$  &$Q^1_{n+1}$ & 0 &$\cdots$ & 0  &$\tilde{Q}^1$ &0 \\
  $U(1)_2$  & 0 &$\cdots$  &0 & $Q^2_1$ &$\cdots$  &$Q^2_{m+1}$  &0 &$\tilde{Q}^2$\\
\bottomrule
\end{tabular}
\caption{Gauge charge table for $\WW\PP^{n|1}\times \WW\PP^{m|1}$.}
\label{tab:productgauge}
\end{table}

This model has two $D$-terms associated with each $U(1)$ respectively,
\begin{equation}
  D_1 \:\sim\: \sum_{i=1}^nQ_i^1|\phi_{1,i}|^2 + \tilde{Q}^1|\psi_1|^2 - r_1\, \quad D_2 \:\sim\: \sum_{i=1}^mQ_i^2|\phi_{2,i}|^2 + \tilde{Q}^2|\psi_2|^2 - r_2\,.
\end{equation}
This is a two-parameter model, and the phase diagram is parameterized by $r_1$ and $r_2$. There are four regions: I $\left\{ r_1 \gg 0,\ r_2 \gg 0\right\}$, II $\left\{ r_1 \ll 0,\ r_2 \ll 0\right\}$, III $\left\{ r_1 \ll 0,\ r_2 \gg 0\right\}$ and IV $\left\{ r_1 \gg 0,\ r_2 \ll 0\right\}$. The phase I, namely, when $\left\{ r_1 \gg 0,\ r_2 \gg 0\right\}$, is the geometric phase, and the vanishing $D$-terms give the product space $\WW\PP^{n|1}\times \WW\PP^{m|1}$.

In phase II, where $ r_1 \ll 0$ and $ r_2 \ll 0$, the two vanishing $D$-terms determine the expectation value for both $\vev{\bpsi_1\psi_1}$ and $\vev{\bpsi_2\psi_2}$. There are residue symmetries, $\ZZ_{\tilde{Q}^1}$ and $\ZZ_{\tilde{Q}^2}$, which keep $\vev{\bpsi_1\psi_1}$ and $\vev{\bpsi_2\psi_2}$ invariant, where $\ZZ_{\tilde{Q}^1}$ acts on $\phi_{1,i}$ and $\psi_1$ while $\ZZ_{\tilde{Q}^2}$ acts on $\phi_{2,i}$ and $\psi_2$. Therefore, this phase is purely a LG point, which is given by the product of two orbifolds: $\left(\CC^n/\ZZ_{\tilde{Q}^1} \right)\times \left(\CC^m/\ZZ_{\tilde{Q}^2}\right)$.

The phase III and IV are two hybrid phases. In this toy example, there are no mixing terms in two $D$-terms, so low energy effective theories on these two phases can be written as the product of an orbifold and an super weighted projective space. For example, In phase III, one could obtain the orbifold $\CC^n/\ZZ_{\tilde{Q}^1}$ from $D_1 = 0$ and the weighted projective superspace $\WW\PP^{m|1}$. Therefore, this phase gives the product $\left(\CC^n/\ZZ_{\tilde{Q}^1}\right) \times \WW\PP^{m|1}$. Similarly, the phase IV is the product $\WW\PP^{n|1} \times \left(\CC^m/\ZZ_{\tilde{Q}^2}\right)$.

\begin{table}[h!]
\centering
\renewcommand{\arraystretch}{1.5}
\begin{tabular}{ccccccccc}
\toprule
  ~  & $\phi_{1,1}$ & $\cdots$ & $\phi_{1,n+1}$ & $\phi_{2,1}$ & $\cdots$ & $\phi_{2,m+1}$ &$\psi_1$ &$\psi_2$ \\\midrule 
  $\left.R\right|_{I}$ & 0  &$\cdots$  &0 & 0  &$\cdots$  &0 &0 &0\\
  $\left.R\right|_{II}$ & $2\frac{Q^1_1}{\tilde{Q}^1}$  &$\cdots$  &$2\frac{Q^1_{n+1}}{\tilde{Q}^1}$ & $2\frac{Q^2_{1}}{\tilde{Q}^2}$  &$\cdots$  &$2\frac{Q^2_{m+1}}{\tilde{Q}^2}$ &$2$ &$2$\\
  $\left.R\right|_{III}$ & $2\frac{Q^1_1}{\tilde{Q}^1}$  &$\cdots$  &$2\frac{Q^1_{n+1}}{\tilde{Q}^1}$ & 0  &$\cdots$  &0 &$2$ &$0$\\
  $\left.R\right|_{IV}$ & 0  &$\cdots$  &0  & $2\frac{Q^2_{1}}{\tilde{Q}^2}$  &$\cdots$  &$2\frac{Q^2_{m+1}}{\tilde{Q}^2}$ &$0$ &$2$\\
\bottomrule
\end{tabular}
\caption{R-charge table for $\WW\PP^{n|1}\times \WW\PP^{m|1}$.}
\label{tab:productR}
\end{table}

The R-charges in the four different phases have been summarized in Table~\ref{tab:productR}. A straightforward generalization of the above discussion applies to the product space of $k$ weighted superspace: $\WW\PP^{n_1|1}\times \WW\PP^{n_2|1} \times \cdots \times \WW\PP^{n_k|1}$. This is also one special case as pointed out in \cite{Belhaj:2004ts}.

\section{Mirror Symmetry}
\label{sec:mirror}
Mirror symmetry for the GLSM for supermanifold was originally proposed by \cite{Aganagic:2004yh} and be further studied by, for example \cite{Belhaj:2004ts,Garavuso:2011nz}. The mirror map is established by the following correspondence:
\begin{align*}
	\Phi_i &\:\longleftrightarrow\: Y_i \, ,\\
	\Psi_\alpha & \:\longleftrightarrow\: X_\alpha,\ \eta_\alpha,\ \chi_\alpha \, ,
\end{align*}
namely, the mirror of a Grassmann odd field $\Psi_\alpha$ is given by a triplet $(X_\alpha, \eta_\alpha, \chi_\alpha)$. The superpotential of the mirror Landau-Ginzburg model is 
\be
\label{eq:mirrorlg}
\begin{aligned}
  W(Y_i,X_\alpha,\eta_\alpha,\chi_\alpha) \:=\:& \sum_i \exp^{-Y_i} + \sum_\alpha \exp^{-X_\alpha}(1 + \eta_\alpha \chi_\alpha) \\
  &\quad + \Sigma \big(\sum_i Q_i Y_i - \sum_\alpha Q_\alpha X_\alpha - t\big)\,,
\end{aligned}
\ee
where the auxiliary field $\Sigma$ serves as the Lagrangian multiplier and it puts the constraint $\sum_i Q_i Y_i - \sum_\alpha Q_\alpha X_\alpha = t $ in the path integral after integrating out:
\be
\label{eq:mirrorlgpf}
\begin{aligned}
    Z \:=\: & \int \big( \prod_{i} \dd Y_i \big)\big( \prod_{\alpha} \dd X_{\alpha} \dd \eta_\alpha \dd \chi_\alpha \big) \delta\big(\sum_i Q_i Y_i - \sum_\alpha Q_\alpha X_\alpha - t\big) \\
    &\quad \times \exp \big( \sum_i \exp^{-Y_i} + \sum_\alpha \exp^{-X_\alpha}(1 + \eta_\alpha \chi_\alpha)\big).
\end{aligned}
\ee
The reasoning for this mirror proposal in \cite{Aganagic:2004yh} was based on central charge consistence. Based on our previous work \cite{Gu:2018xzx}, in this section we want to give more evidence for this mirror proposal by matching chiral ring relations and correction functions. 

For a general purpose, we will start with the more general topological (B-twisted) super Landau-Ginzburg model and then perform the computations in the general setup, such as the correlation functions given in . We will see the mirror Landau-Ginzburg model (\ref{eq:mirrorlg}) can be viewed as one special case of our general definitions, so the general results can be applied to the mirror Landau-Ginzburg model. For the A-twisted GLSM for supermanifolds, we have given the formula for correlation functions by supersymmetric localization in \cite{Gu:2018xzx}. By directly comparing the correlation functions from \cite{Gu:2018xzx} and  Section~\ref{sec:corf}, we claim that the mirror symmetry for supermanifolds is verified.

\begin{figure}[!h]
\centering
\begin{tikzpicture}[every text node part/.style={align=center},>=angle 90]
  \draw (0,0) node (n1) {GLSM \\ for supermanifold};
  \draw (0,-3.5) node (n2) {GLSM \\ for hypersurface};
  \path[<->, line width=0.3mm, font = \footnotesize ](n2) edge node[draw=none,sloped, anchor=center, above] {Equivalent} (n1);
  \draw (6,0) node (n3) {Mirror \\ super-LG};
  \draw (6,-3.5) node (n4) {Mirror LG};
  \path[<-,line width=0.3mm] (n4) edge node[draw=none,sloped, anchor=center, above]{Integrate \\  out $\eta_i, \chi_i$} (n3);
  \path[<->,line width=0.3mm] (n1) edge node[draw=none,sloped, anchor=center, above]{Mirror} (n3);
  \path[<->,line width=0.3mm] (n2) edge node[draw=none,sloped, anchor=center, above]{Mirror} (n4);
\end{tikzpicture} 
\caption{Relations between GLSM for a supermanifold and for the corresponding hypersurface and their mirrors.}
\label{fig:relation2}
\end{figure}
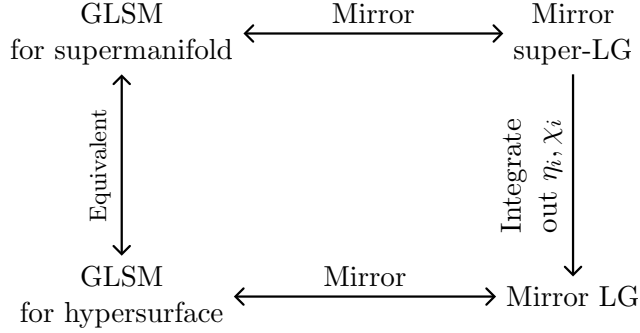

In \cite{Aganagic:2004yh,Garavuso:2011nz}, they have used the mirror model to check the equivalence statement in \cite{Sethi:1994ch,Schwarz:1995ak}. By integrating out the odd chiral superfields, $\eta_\alpha$ and $\chi_\alpha$, in (\ref{eq:mirrorlgpf}), it can be shown that the remain part is just the mirror Landau-Ginzburg model for a GLSM for a hypersurface (or complete intersection). Their relations can be described by the diagram Figure~\ref{fig:relation2}. However, in this picture, the mirror symmetry for supermanifolds need to be checked. This is the motivation for this section. 

\subsection{Topological Super Landau-Ginzburg Model}
\label{sec:tlg}

Before going to the mirror Landau-Ginzburg model, we shall start with the general topological super Landau-Ginzburg model. Our definition of topological super Landau-Ginzburg model generalize to the definitions in \cite{Vafa:1990mu,Guffin:2008kt}. More specifically, it is B-twisted super Landau Ginzburg. The definition is quite straight forward by including Grassmannian odd chiral fields. The action can be written as
\begin{equation}
\label{eq:genermodel}
	S \:=\: \int {\rm d}^2 z {\rm d}^4 \theta K(X^I, \bar{X}^{\bar{I}})  + \int {\rm d}^2 z {\rm d}^2\theta W(X^I) + c.c.,
\end{equation}
where index $I$ runs $\{i,\mu\}$ with $i$ indicates even chiral field while $\mu$ indicates odd chiral field. Consider the canonical kinetic terms, {\it i.e.} $K(X^I, \bar{X}^{\bar{I}}) = \sum_I X^I \bar{X}^{\bar{I}}$, then the action written in components \footnote{In the following cases where no confusion arises, we may interchangeably use the same notation for the superfield and its scalar component.} is:
\begin{equation}
\label{eq:canmodel}
	S \:=\: \int{\rm d}^2 z \left[|\partial X^I|^2 + |\partial_I W|^2 + \rho^I\bar{\partial}\psi^{\bar{I}} + \bar{\rho}^I \partial \bar{\psi}^{\bar{I}} + \rho^I \partial_I\partial_JW \bar{\rho}^J + \bar{\psi}^{\bar{I}} \partial_{\bar{I}} \partial_{\bar{J}}W \psi^{\bar{J}}  \right]
\end{equation}
Then it is invariant under following SUSY-transformation or BRST-like transformation is
\begin{align*}
	&\delta X^I \:=\: 0, &&\delta \bar{\psi}^{\bar{I}} \:=\: -\partial_{I} W,  &&\delta \rho^I \:=\: -\partial X^I,\\
	&\delta X^{\bar{I}} \:=\: \psi^{\bar{I}} + \bar{\psi}^{\bar{I}}, &&\delta \psi^{\bar{I}} \:=\: \partial_{I}W, &&\delta \bar{\rho}^{I} \:=\: - \bar{\partial} X^I.
\end{align*}
Note that $X^I$ transformation is trivial and then the physical observables must be functions of $X^I$'s. Following \cite{Cecotti:1991me}, the chiral ring relation can be obtained from the equations of motion. The only difference is that we have two sets of equations of motion for even chiral fields and odd chiral fields. Firstly, take the variation with respect to even chiral fields $X^i$:
\begin{equation*}
	\partial_i W(X^I) \:\sim\: D^+\bar{D}^{+}\partial_i K(X^I, \bar{X}^{\bar{I}})
\end{equation*}
On the left hand side, it is a Q-closed operator. Therefore, $\partial_i W(X^I) = 0$ shall give part of the chiral ring relations. Now, perform the variation with respect to odd chiral fields $X^{\mu}$. Here, we make the following assumptions:
\begin{enumerate}
	\item The \kahler potential and the superpotential are Grassmannian even,
	\item A function of even variables cannot be odd.
\end{enumerate}
Then, the \kahler  potential expands in terms of these odd chiral fields up to finite terms, and the expansion in terms of odd chiral fields would be:
\begin{align*}
	K(X^I,\bar{X}^{\bar{I}}) \:=\:& K_0(X^i,\bar{X}^{\bar{i}}) + X^{\mu}X^{\nu} K_{\mu\nu}(X^i,\bar{X}^{\bar{i}}) + X^\mu \bar{X}^{\bar{\nu}} K_{\mu\bar{\nu}}(X^i, \bar{X}^{\bar{i}}) \\
	&+ \bar{X}^\mu \bar{X}^{\bar{\nu}} K_{\bar{\mu}\bar{\nu}}(X^i, \bar{X}^{\bar{i}}) +X^{\mu}X^{\nu}X^{\rho}X^{\lambda} K_{\mu\nu\rho\lambda}(X^i,\bar{X}^{\bar{i}})+ \cdots,\\
	W(X^I) \:=\:& W_0(X^i) + X^{\mu}X^{\nu} W_{\mu\nu}(X^i) + X^{\mu}X^{\nu}X^{\rho}X^{\lambda} W_{\mu\nu\rho\lambda}(X^i) + \cdots.
\end{align*}
In the above, the coefficient functions are even functions and are anti-symmetric in indices. Then the equation of motion for the odd chiral fields can be read as,
\begin{align*}
	\partial_{\mu}W(X^i) &\:\equiv\:  X^{\nu} W_{\mu\nu}(X^i) + X^{\nu}X^{\rho} W_{\mu\nu\rho}(X^i)+\cdots \\
	&\:\sim\: D^+\bar{D}^{+}\left( X^{\nu} K_{\mu\nu}(X^i,\bar{X}^{\bar{i}})+ \cdots\right)
\end{align*}
Therefore, $\partial_{\mu}W(X^i)=0$ shall also contribute to the chiral ring relations. Putting them together $\partial_I W(X) = 0$ for $I={i,\mu}$, the full chiral ring can be written as
\begin{equation*}
	\mathcal{R} \:=\: \frac{\mathbb{C}[X]}{\partial_I W(X)}\,.
\end{equation*}

As we can see from the above, the mirror proposal in \cite{Aganagic:2004yh} is a special case where there are two odd chiral fields $(\eta, \chi)$ with a specific choice of the superpotential.

\subsection{Correlation functions}
\label{sec:corf}
Starting with (\ref{eq:canmodel}) as a CFT, in which case the superpotential should be quasi-homogeneous, we can rescale the world-sheet metric and take $g\rightarrow \lambda^2 g$ \cite{Vafa:1990mu,Guffin:2008kt}:
\begin{equation*}
	S \:=\: \int{\rm d}^2 z \left[|\partial X^I|^2 + \lambda^2|\partial_I W|^2 + \rho^I\bar{\partial}\psi^{\bar{I}} + \bar{\rho}^I \partial \bar{\psi}^{\bar{I}} + \rho^I \partial_I\partial_JW \bar{\rho}^J + \lambda^2 \bar{\psi}^{\bar{I}} \partial_{\bar{I}} \partial_{\bar{J}}W \psi^{\bar{J}}  \right].
\end{equation*}
Take large $\lambda$ limit, the path integral is dominated near critical points determined by:
\begin{align*}
	X^I &\:=\: 0\,, \\
	\partial_J W(X) &\:=\: 0\,.
\end{align*}
In this limit, the path integral is localized to these critical points and can be calculated exactly to be the ratio of determinants of odd chiral fields and even chiral fields \cite{Vafa:1990mu,Guffin:2008kt}:
\be
	Z_{\rm total}^g \:=\: \left(\frac{H_{\rm even}}{H_{\rm odd}} \right)^{g-1}\,,
\ee
where 
\be
	H_{\rm even} \:=\: \det(\partial_i \partial_j W)|_{\rm crit.\ pts.}, \quad H_{\rm odd} \:=\: \det(\partial_{\mu}\partial_{\nu}W)|_{\rm crit.\ pts.},
\ee
and $g$ the genus of the world-sheet. More details can be found in Appendix \ref{app:corf}. Therefore, the general correlation function is
\be
\label{eq:lgcor}
	\vev{\cO_1(X)\cO_2(X)\cdots \cO_N(X)} \:=\: \sum_{\{X_c | \dd W = 0\}}Z_{\rm total}^g \cO_1(X_c) \cO_2(X_c) \cdots \cO_N(X_c).
\ee
In principle, above general operators are functions of $X_I$'s. 

A remark should be added is that here we have only considered the correlation functions in super topological Landau-Ginzburg models, which have no orbifolds. It would be interesting to extend the current discussion to the orbifold cases, see, for example, \cite{Gu:2020nub} for the ordinary Landau-Ginzburg model with $\ZZ_2$-orbifold.

\subsection{Verification of mirror proposal}
\label{sec:mirrorverif}

There are several ways of checking the mirror symmetry following \cite{hori2003mirror,Gu:2017nye,Gu:2018fpm}. Here, we will check the mirror symmetry for supermanifolds by verifying that some BPS quantities are consistent, namely chiral ring relations and correlation functions. In this subsection, we only verify the example $\WW\PP^{4|1}$ when $g=0$ case. The higher genus cases can be obtained by similar calculations. The argument presented in this section can be generalized to other examples straightforwardly.

Let us recall the mirror Landau-Ginzburg for the GLSM for $\WW\PP^{4|1}$. It is defined by
\begin{itemize}
  \item $5$ even chiral superfields, $Y_i$,
  \item $1$ even chiral superfields, $X$,
  \item $2$ odd chiral superfields, $\eta$ and $\chi$,
  \item the auxiliary field $\Sigma$, which is even,
\end{itemize}
with superpotential
\be
\label{eq:superpotentialex}
  W \:=\: \Sigma\left( \sum_{i=1}^5 Q_i Y_i - \tilde{Q} X - t \right) + \sum_{i=1}^5 e^{-Y_i} + e^{-X} (1+ \eta \chi)\, .
\ee

\paragraph{Chiral ring relations.} First, note that integrating out $\Sigma$ gives the relation (we shall remark again that we will interchangeably use the same notation for the superfield and its scalar component)
\be
  \sum_{i=1}^5 Q_i Y_i - \tilde{Q} X - t \:=\: 0\, .
\ee
Without loss of the generality, we assume $Q_5 = 1$ and then
\be
  Y_5 \:=\: t + \tilde{Q}X - \sum_{i=1}^4 Q_i Y_i\, .
\ee
By the redefinition
\be
  \Pi \equiv \exp(-Y_5) \:=\: q e^{- \tilde{Q}X} \prod_{i=1}^4 e^{ Q_i Y_i}\,,
\ee
the superpotential becomes
\be
  W \:=\: \sum_{i=1}^4 e^{-Y_i} + \Pi + e^{-X} (1+\eta\chi)\,.
\ee
On the critic loci, we shall have
\be
\begin{aligned}
  \frac{\partial W}{\partial Y_i} &\:=\: - e^{-Y_i} + Q_i \Pi \:=\: 0 \,, \\
  \frac{\partial W}{\partial X} &\:=\: - \tilde{Q}\Pi - e^{-X}(1+\eta\chi) \:=\: 0 \,, \\
  \frac{\partial W}{\partial \eta} &\:=\: e^{-X} \chi  \:=\: 0 \,, \\
  \frac{\partial W}{\partial \chi} &\:=\: - e^{-X} \eta  \:=\: 0 \,. 
\end{aligned}
\ee
Therefore,
\be
  e^{-Y_i} \:=\: Q_i \Pi\,,\quad e^{-X} \:=\: - \tilde{Q}\Pi,\, \quad \eta \:=\: \chi \:=\: 0\,. 
\ee
Then the chiral ring relation can be read as
\be
\label{eq:chiralring}
  \left(\prod_{i=1}^5(Q_i \Pi)^Q_i \right)\left(-\tilde{Q}\Pi\right)^{-\tilde{Q}} \:=\: q\,.
\ee
The above relation is the same as the chiral ring relation obtained from the Landau-Ginzburg model mirror to the GLSM for the hypersurface of degree $\tilde{Q}$ in $\WW\PP^4$. By the identification \footnote{For the general abelian $U(1)^k$ case, this map should be $Q_i^a\Pi_a \:\leftrightarrow\: Q_i^a\sigma_a$.} $$\Pi\:\longleftrightarrow\: \sigma,$$ where the $\sigma$ on the right hand side corresponds to the scalar component of the $U(1)$ gauge field strength in the mirror GLSM,  the above ring relation \eqref{eq:chiralring} is the same as the chiral ring relation obtained in the GLSM for $\WW\PP^{4|1}$, up to an factor of $(-1)^{\tilde{Q}}$ \cite{Gu:2018xzx}.

\paragraph{Correlation functions.} In this following, we directly apply the formula (\ref{eq:lgcor}) to calculate correlation functions for this example. Starting with the superpotential \eqref{eq:superpotentialex}, the critical loci are determined by $dW = 0$, leading to the following equations
\begin{gather*}
  \sum_{i} Q_i Y_i - \tilde{Q} X - t \:=\: 0, \\
  - \Sigma \tilde{Q} \:=\: e^{-X}, \\
  \Sigma Q_i \:=\: e^{-Y_i},\\
  \chi \:=\: \eta \:=\: 0. 
\end{gather*}
The Hessian matrix can be dived into four blocks
\be
\begin{bmatrix}
  A &B \\
  C &D \\
\end{bmatrix}_{dW=0},
\ee
where $A$ is the even part:
\be
A \:=\: 
\setlength{\arraycolsep}{2.5pt}
\renewcommand\arraystretch{1.2}
\begin{bmatrix}
  \frac{\partial^2 W}{\partial \Sigma^2} &\frac{\partial^2 W}{\partial \Sigma\partial X } &  \frac{\partial^2 W}{\partial \Sigma\partial Y_i} \\
  \frac{\partial^2 W}{\partial X \partial \Sigma } &  \frac{\partial^2 W}{\partial X^2  } & \frac{\partial^2 W}{\partial X \partial Y_i } \\
  \frac{\partial^2 W}{\partial Y_j \partial \Sigma } &  \frac{\partial^2 W}{\partial Y_j \partial X } & \frac{\partial^2 W}{\partial Y_j \partial Y_i }
\end{bmatrix}_{dW=0}
\:=\:
\begin{bmatrix}{ccc}
  0           & -\tilde{Q}        & Q_i \\
  -\tilde{Q}  & -\Sigma\tilde{Q}  & 0  \\
  Q_j         & 0                 & \delta^{ij}\Sigma Q_i
\end{bmatrix}, \no
\ee
$D$ is the odd part:
\be
D \:=\: 
\left[ 
\setlength{\arraycolsep}{2.5pt}
\renewcommand\arraystretch{1.2}
\begin{array}{cc}
  0 & \frac{\partial^2 W }{\partial \eta \partial \chi} \\
  \frac{\partial^2 W }{\partial \chi \partial \eta} & 0 
\end{array}
\right]_{dW= 0} 
\:=\: 
\left[  
\begin{array}{cc}
  0 & -\Sigma \tilde{Q} \\
  \Sigma \tilde{Q}  & 0
\end{array}
\right], \no
\ee
and $B$ and $C$ are the mixed parts which will be zeros on the critical loci. Therefore, direction computations give 
\begin{align}
 H_{\rm even} &\:=\: \det A \:=\: \Sigma \tilde{Q} \left(\prod_{i=1}^5 \Sigma Q_i \right) \left( \sum_{i=1}^5\frac{Q_i}{\Sigma}- \frac{\tilde{Q}}{\Sigma} \right), \\
 H_{\rm odd} &\:=\: \det D \:=\: (\Sigma\tilde{Q})^2,
\end{align}
and so
\be
  H \:=\: \frac{H_{\rm even}}{H_{\rm odd}} \:=\: \frac{\prod_{i=1}^5 \Sigma Q_i}{\Sigma \tilde{Q} } \left( \sum_{i=1}^5\frac{Q_i}{\Sigma}- \frac{\tilde{Q}}{\Sigma} \right)
\ee
Then in the genus zero case, the general correlation function for operator $\cO$ is 
\be
  \vev{\cO} \:=\: \sum_{\{X_c | \dd W = 0\}} \cO \left(\frac{\prod_{i=1}^5 \Sigma Q_i}{\Sigma \tilde{Q} }\right)^{-1} \left( \sum_{i=1}^5\frac{Q_i}{\Sigma}- \frac{\tilde{Q}}{\Sigma} \right)^{-1}.
\ee
Comparing the above correlation function with the result from the A-twisted model \cite{Closset:2015rna,Gu:2017nye,Gu:2018xzx}, the results are the same up to a constant factor if we identify the first bracket in the above as the contribution from one-loop determinant and the second bracket as the Hessian of the twisted one-loop effective action. Above derivation can be generalized to more general cases with multiple $U(1)$'s. We have included more details in Appendix \ref{app:verf}.

As shown in \cite{Aganagic:2004yh,Garavuso:2011nz}, in the mirror Landau-Ginzburg model for a supermanifold, there would be a measure pop out in the path integral after integrating out the odd chiral fields, which is the same as the measure for the fields mirror to $P$-fields. Thereby, from the discussion above as well as from Appendix~\ref{app:verf}, we also further confirm this point that the odd chiral fields have very similar effects as those $P$-fields. This demonstrates the downward arrow shown in Figure~\ref{fig:relation2}.

\section{Outlook}

There are several natural directions to pursue. A nonabelian generalization of GLSMs for supermanifolds is still lack of investigation. It would be interesting to find out what new phenomenon could happen in the nonabelian GLSMs for supermanifolds, and make comparisons with ordinary nonabelian GLSMs, for example \cite{Hori:2006dk,Hori:2011pd}, as well as the nonabelian mirror symmetry \cite{Gu:2018fpm,Gu:2019zkw}. Extending the present analysis to hemisphere/boundary GLSMs \cite{Herbst:2008jq,Hori:2013ika} and to matrix factorizations for super Landau–Ginzburg models, generalizing the work \cite{Kapustin:2003ga,Hori:2004ja,Hori:2004zd}, could clarify D-brane realizations and open-string mirrors. Finally, a careful analysis of fixed loci and orbifold singularities at the LG point (and their effect on higher-genus amplitudes) would strengthen the physical interpretation of the conjectured orbifold description. We will leave these questions for future direction.

\section*{Acknowledgment}
The author would like to thank Wei Gu for many useful discussions and comments for the manuscript. H.Z. was partially supported by the National Natural Science Foundation of China (Grant No.~12405083,~12475005) and the Shanghai Pujiang Program (Grant No.~24PJA119).

\appendix
\section{Partition functions for topological super Landau-Ginzburg models}
\label{app:corf}

Following the set-up in Section \ref{sec:tlg}, the partition function $Z_{\rm total}$ can be divided as the product of the even part $Z_{\rm even}$ and the odd part $Z_{\rm odd}$. The even part is exactly the same as in \cite{Vafa:1990mu,Guffin:2008kt} and is given as:
\begin{equation}
  Z_{\rm even} \:=\: H_{\rm even}^{g-1},
\end{equation}
with $H_{\rm even} = \det(\partial_i\partial_j W)$ and $g$ the genus of worldsheet. Now let us look at the odd part. First, assume we have $n$ odd supermultiplets $X^{\mu}$ and their components are given as $X^{\mu} = (x^{\mu},\rho^{\mu},\psi^{\bar{\mu}},\cdots)$, with $x^\mu$ are bosonic but anti-commutative. The odd part of the partition function can be calculated as following.
\begin{equation}
\begin{aligned}
    Z_{odd,x} \:=\:& \int \prod \dd^2 x^{\mu} \exp\big( -|\lambda \partial_\mu W|^2 \big)  \\
              \:=\:& \int \prod \dd^2 x^{\mu} (\lambda^2)^n x^{\alpha}\bar{x}^{\beta} \partial_\alpha \partial_\mu W \partial_\beta \partial_\nu W  \\
              \:=\:& \lambda^{2n} H_{\rm odd} \bar{H}_{\rm odd},
\end{aligned}
\end{equation}
\begin{equation}
\begin{aligned}
    Z_{odd,\rho,\chi} \:=\:& \left|\int \prod \dd^2\rho \exp\big( -\rho^\mu \partial_\mu  \partial_\nu W  \bar{\rho}^\nu\big)\right|^g  \\
              & \quad \times \int \prod \dd^2\psi \exp\left( -\lambda^2 \bar{\psi}^{\bar{\mu}}\partial_{\bar{\mu}} \partial_{\bar{\nu}} \overline{W} \psi^{\bar{\nu}}\right) \\
              \:=\:& \lambda^{-2n} H_{\rm odd}^{-g} \bar{H}_{\rm odd}^{-1},
\end{aligned}
\end{equation}
with $H_{\rm odd} = \det(\partial_\mu \partial_\nu W)$. Therefore, we shall have
\begin{equation}
    Z_{\rm total} \:=\: Z_{\rm even} Z_{\rm odd} = \left(\frac{H_{\rm even}}{H_{\rm odd}} \right)^{g-1}.
\end{equation}



\section{Verification of mirror proposal for general cases}
\label{app:verf}
In this appendix, we give the calculation for more general cases. We consider $U(1)^k$ gauge with $m$ even chiral superfields $\Phi_i$ with charges $Q_i^a$ and $n$ odd chiral superfields $\tilde{\Phi}_\mu$ with charges $\tilde{Q}_{\mu}^a$. The mirror Landau-Ginzburg model is defined by following data:
\begin{itemize}
  \item $m$ even chiral superfields, $Y_i$,
  \item $n$ even chiral superfields, $X_\mu$,
  \item $n$ pair of odd chiral superfields, $\eta_\mu$ and $\chi_\mu$,
  \item auxiliary fields $\Sigma_a$,
\end{itemize}
with superpotential
\be
  W \:=\: \sum_{a=1}^k \Sigma_a \left( \sum_{i=1}^m Q_i^a Y_i - \sum_{\mu=1}^n\tilde{Q}_\mu^a X_{\mu} -t^a\right) + \sum_{i=1}^m e^{-Y_i} + \sum_{\mu=1}^n e^{-X_\mu} (1+ \eta_{\mu}\chi_{\mu})\,.
\ee
The critical loci are determined by $dW = 0$ as below
\begin{gather*}
  t^a \:=\: \sum_{i=1}^m Q_i^a Y_i - \sum_{\mu=1}^n\tilde{Q}_\mu^a X_{\mu}\,, \\
  Q_i^a\Sigma_a \:=\: e^{-Y_i}\,, \\
  -\tilde{Q}_{\mu}^a\Sigma_a \:=\: e^{-X_{\mu}}\,, \\
  \eta_{\mu} \:=\: \chi_{\mu} = 0\,.
\end{gather*}
The Hessian matrix for this superpotential also includes: even part, odd part and mixed parts. The mixed parts are zero on the critical loci. The even part can be written as:
\be
\setlength{\arraycolsep}{2.5pt}
\renewcommand\arraystretch{1.3}
\begin{bmatrix}
  \frac{\partial^2 W}{\partial \Sigma_a\Sigma_b} &\frac{\partial^2 W}{\partial \Sigma_a \partial X_\mu } &  \frac{\partial^2 W}{\partial \Sigma_a \partial Y_i} \\
  \frac{\partial^2 W}{\partial X_\nu \partial \Sigma_b } &  \frac{\partial^2 W}{\partial X_\nu \partial X_\mu  } & \frac{\partial^2 W}{\partial X_\nu \partial Y_i } \\
  \frac{\partial^2 W}{\partial Y_j \partial \Sigma_b } &  \frac{\partial^2 W}{\partial Y_j \partial X_\mu } & \frac{\partial^2 W}{\partial Y_j \partial Y_i }
\end{bmatrix}_{dW=0}
\:=\: 
\renewcommand\arraystretch{1.2}
\begin{bmatrix}
  0                   & -\tilde{Q}_\mu^a  & Q_i^a \\
  -\tilde{Q}_{\nu}^b  & -\delta^{\mu\nu}\displaystyle\sum_{c} \tilde{Q}_\nu^c\Sigma_c  & 0  \\
  Q_j^b         & 0                 & \delta^{ij} \displaystyle\sum_{c} Q_i^c \Sigma_c
\end{bmatrix}\,, \no
\ee
The odd part can be written as:
\be 
\begin{bmatrix}
  0 & \frac{\partial^2 W }{\partial \eta_{\mu} \partial \chi_\nu} \\
  \frac{\partial^2 W }{\partial \chi_{\mu} \partial \eta_\nu} & 0 
\end{bmatrix}_{dW= 0} 
\:=\:   
\setlength{\arraycolsep}{2.5pt}
\renewcommand\arraystretch{1.2}
\begin{bmatrix}
  0 & - \delta^{\mu\nu} \displaystyle\sum_{c} \tilde{Q}_\nu^c\Sigma_c \\
  \delta^{\mu\nu}\displaystyle\sum_{c} \tilde{Q}_\mu^c\Sigma_c & 0
\end{bmatrix}\,. \no
\ee
Therefore, we have
\begin{align*}
  H_{\rm even} &\:=\: \left( \prod_{\mu}\tilde{Q}_{\mu}^a\Sigma_a \right) \left( \prod_{i}Q_i^a\Sigma_a \right) \det\left( \sum_i\frac{Q_i^a Q_i^b}{Q_i^c\Sigma_c} - \sum_{\mu} \frac{\tilde{Q}_{\mu}^a\tilde{Q}_{\mu}^b}{\tilde{Q}_{\mu}^c\Sigma_c}\right), \\
  H_{\rm odd} &\:=\: \left(\prod_{\nu} \tilde{Q}_\nu^a\Sigma_a\right)^2,
\end{align*}
and
\be
  H \:=\: \frac{H_{\rm even}}{H_{\rm odd}}\:=\: \left(\frac{\prod_{i}Q_i^a\Sigma_a }{\prod_{\mu}\tilde{Q}_{\mu}^a\Sigma_a}\right) \det\left( \sum_i\frac{Q_i^a Q_i^b}{Q_i^c\Sigma_c} - \sum_{\mu} \frac{\tilde{Q}_{\mu}^a\tilde{Q}_{\mu}^b}{\tilde{Q}_{\mu}^c\Sigma_c}\right)\,.
\ee
Compared with known results from A-twisted model calculations \cite{Closset:2015rna,Gu:2017nye,Gu:2018xzx}, the first bracket will correspond to the one-loop determinant from chiral fields and the second determinant will correspond to the twisted one-loop effective action, which is from the Hessian of the effective twisted superpotential.

\bibliographystyle{JHEP}
\bibliography{ref}

\end{document}